\documentclass[12pt]{article}
\usepackage{graphicx}
\usepackage{epstopdf}
\usepackage{geometry}
\usepackage{url}
\title{ Keyspace: A Consistently Replicated, Highly-Available Key-Value Store }
\author{ Marton Trencseni, \texttt{mtrencseni@gmail.com} \and Attila Gazso, \texttt{agazso@gmail.com} }

\begin{document}

\maketitle

\sloppy
\abstract{ This paper describes the design and architecture of Keyspace, a distributed key-value store offering strong consistency, fault-tolerance and high availability. The source code is available under the open-source AGPL license for Linux, Windows and BSD-like platforms. As of 2012, Keyspace is no longer undergoing active development. }

\section{ Introduction }

This paper describes the design and architecture of Keyspace, a distributed key-value store, offering replication, fault-tolerance, high availability and consistency. Below we define what these terms mean in the context of Keyspace:

\begin{enumerate}
\item Key-value store: a database offering only basic operations such as \texttt{SET(key, value)} and \texttt{GET(key)}. For a complete list see section 'Operations'.
\item Distributed: several computers (nodes) store data and state, exchange messages and perform computation to serve client requests.
\item Consistent: all clients see the database going through the same sequence of states (key-value pairs). This means clients do not have to worry about different parts of the distributed system being out of sync.
\item Replicated: all nodes store the same data, in this case key-value pairs and database state. The storage capacity of the distributed system is limited by the storage capacity of the individual nodes.
\item Fault-tolerant: the system can tolerate certain failure conditions and continue serving client requests. Keyspace can tolerate a minority of nodes failing, ie. it will continue to serve client requests if a majority of nodes are alive and connected.
\item Highly-available: the system continuously serves client request. In the case of Keyspace, replication and fault-tolerance guarantee high availability.
\end{enumerate}

The intended use of Keyspace is as a building block for other distributed systems, ie. the lowest layer in a distributed stack, for example as a controller which distributes data and workload across other machines. Alternatively, Keyspace may be used in itself to store and serve small-to-medium amounts of data in a scenario when replication and high-availability are desired. Keyspace is released under the open-source AGPL \cite{AGPL} license for Linux, Windows, and BSD-like operating systems (eg. MacOS X), the source code is available at \url{https://github.com/scalien/keyspace}.

\section{ Consistency }

Keyspace makes strong consistency guarantees by using the Paxos distributed consensus algorithm \cite{PaxosMadeSimple}. The advantage of strong consistency is that Keyspace acts as a "regular" key-value store, meaning clients need not bother with distributed aspects. Consistency, in the context of Keyspace is the guarantee that once a write operation completes:

\begin{enumerate}
\item The data has been written to disk on a \textit{majority} of nodes.
\item All succeeding read operations will reflect the data written.
\end{enumerate}

The trade-off for strong consistency is that Keyspace will only complete write operations if a majority of nodes are alive and able to communicate. This can be amortized by increasing the number of nodes forming the Keyspace system. The table below shows the probability of liveness for different system sizes assuming the probability of a node being alive is 95\%.

\begin{table}[h]
\begin{center}
\begin{tabular}{ | c | c | c | }
\hline
total nodes & required for majority & probability of liveness \\
\hline \hline
1 & 1 & 95.00\% \\ \hline
2 & 2 & 90.25\% \\ \hline
\textbf{3} & \textbf{2} & \textbf{99.27\%} \\ \hline
4 & 3 & 98.59\% \\ \hline
5 & 3 & 99.88\% \\ \hline
.. & .. & .. \\ \hline
\end{tabular}
\caption{ Probability of liveness for different system sizes. }
\end{center}
\end{table}

Clients can choose to trade consistency for availability and issue special dirty-read requests, which are always returned by a single node. Dirty-reads do not guarantee consistency, ie. they may not reflect all previous write operations the Keyspace system has performed. (Sometimes regular reads are referred to as safe-reads to differentiate.) There is no dirty-write operation --- for Keyspace to perform a write operation, a majority of nodes have to be alive.

Paxos, in the context of Keyspace is used to replicate database write commands. All nodes receive these commands in the same order and execute them on their local database. This guarantees that all nodes' local databases go through the same sequence of states, hence the name replicated database.

\section{ Fault-tolerance }

Keyspace guarantees consistency under all failure conditions occuring in real world computing networks such as data centers or the Internet. These failure conditions are:

\begin{enumerate}
\item Nodes stop and restart: the computers running the Keyspace program may stop and restart, losing in-memory state but not losing data written to disk.
\item Network splits: switches and other networking equipment may fail, causing the nodes to split into disjoint networks.
\item Message loss, duplication and reordering: operating system network stacks and routers may lose or reorder messages. Network protocols such as TCP guarantee reliable delivery in sent order, while UDP does not. Keyspace can run over both TCP and UDP-like protocols.
\item In-transit message delays: on busy networks and WAN environments such as the Internet, messages may take several seconds to arrive at the recipient.
\end{enumerate}

Keyspace keeps its strong consistency guarantees, as described before, under all enumerated fault conditions. This means that if enough of the Keyspace system is functioning (a majority of nodes are up and connected without too much packet loss) then the users see the distributed key-value store which behaves like a "regular" key-value store.

To break the consistent replication of Keyspace requires disk failure corrupting the locally stored database or an administrator accidentally deleting it altogether.

\section{ Liveness }

Keyspace is a distributed system that requires a majority of nodes to be alive and connected to safely serve client requests. Table 1. showed how many nodes are required for majority for different system sizes, with the configuration recommended for most users highlighted: users wishing to tolerate single-machine failure should use $n=3$. Users looking for even more fault-tolerance should use $n=5$ guaranteeing liveness in case of two machine failures.

Notice that the probabilities for even sized systems are lower than for odd sized systems, eg. the probability of liveness in the example is 99.27\% for $n=3$ and 98.59\% for $n=4$, meaning that adding a machine (the fourth) to the system \textit{decreases} the probability of liveness. The explanation is intuitive: in both setups only one machine can fail, but in the $n=4$ case three machines have to be up, whereas in the $n=3$ case only two. The conclusion is that users are advised to use odd sized Keyspace systems: $n=3, 5, 7...$

\section{ Operations }

Keyspace supports the following client commands (note the dirty-read operations at the end of the list):

\begin{itemize}
\sloppy
\item \texttt{GET(key)}: returns the value of \texttt{key}, if it exists in the database.
\item \texttt{SET(key, value)}: sets the value of \texttt{key}, overwriting the previous value if it already existed in the database.
\item \texttt{TEST-AND-SET(key, test, value)}: atomically changes the value of \texttt{key} to \texttt{value} if its current value is \texttt{test}.
\item \texttt{ADD(key, a)}: treats the value of \texttt{key} as a number and atomically increments it by \texttt{a}, where \texttt{a} may be a negative number. Useful for building indexes and counters.
\item \texttt{RENAME(key, newKey)}: renames \texttt{key} to \texttt{newKey}.
\item \texttt{DELETE(key)}: deletes \texttt{key} and its value from the database.
\item \texttt{REMOVE(key)}: deletes \texttt{key} and its value from the database and returns the value.
\item \texttt{PRUNE(prefix)}: deletes all key-value pairs where key starts with \texttt{prefix}.
\item \texttt{LIST-KEYS(prefix, startKey, count, next, forward)}: returns at most \texttt{count} keys whose beginning matches \texttt{prefix}, starting at position \texttt{startKey} in the database. If \texttt{startKey} is not found in the database, the listing starts at the next item lexicographically. If \texttt{startKey} is found, \texttt{startKey} itself may be skipped by passing in \texttt{next = true}. This is useful for pagination, where you pass in the last key that you showed on the previous page, but don't want to see it at the top of the next page. \texttt{forward} specifies the direction of the listing.
\item \texttt{LIST-KEYVALUES(prefix, startKey, count, next, forward)}: same as \texttt{LIST-KEYS}, but returns keys \textit{and values}.
\item \texttt{COUNT(prefix, startKey, count, next, forward, forward)}: returns the number of items the same \texttt{LIST} operation with the same arguments would return.
\item \texttt{DIRTY-GET(key)}: returns the value of \texttt{key}, if it exists in the database, without making any consistency guarantees about reflecting previous write operations.
\item \texttt{DIRTY-LIST-KEYS(prefix, startKey, count, next, forward)}: same as \texttt{LIST-KEYS}, but without making any consistency guarantees about reflecting previous write operations.
\item \texttt{DIRTY-LIST-KEYVALUES(prefix, startKey, count, next, forward)}: same as \newline \texttt{LIST-KEY-VALUES}, but without making any consistency guarantees about reflecting previous write operations.
\item \texttt{DIRTY-COUNT(prefix, startKey, count, next, forward)}: same as \newline \texttt{COUNT}, but without making any consistency guarantees about reflecting previous write operations.
\end{itemize}

\section{ Master leases }

Keyspace is a master-slave system. Any node can obtain a master lease, which is valid for a maximum of 7 seconds, and can extend the lease before it expires, thus holding on to its mastership as long as it doesn't fail and network conditions are favorable. If the master fails, another node (previously a slave) will take over mastership within a few seconds. It is guaranteed that this process is safe, ie. only a single node will believe itself to be the master.

In Keyspace, only the master node serves read and write operations. but all nodes will perform dirty-reads. The advantage is that, since all write operations go through the master, it can be sure that is has seen all write operations, thus it can safely serve read operations without contacting other nodes. This means read operations are cheap, while write operations are not more expensive. The downside is that only a single node services client requests, meaning it could become a performance bottleneck.

Keyspace uses the PaxosLease protocol \cite{PaxosLease} for master leases, a variant of Paxos which does not require disk writes during lease negotiation. PaxosLease does not require clock synchrony between the nodes participating in the distributed algorithm.

\section{ Architecture }

Keyspace was designed for strong consistency from the ground-up. The basic distributed primitive is Leslie Lamport's Paxos consensus algorithm. \textbf{Paxos} relies on a message transport layer which it uses to send messages to other nodes. Paxos does not assume reliable message delivery: messages can get lost, reordered or delayed. These fault-conditions are handled by the Paxos algorithm itself.

\begin{figure}[htbp]
\begin{center}
\includegraphics[scale=0.5]{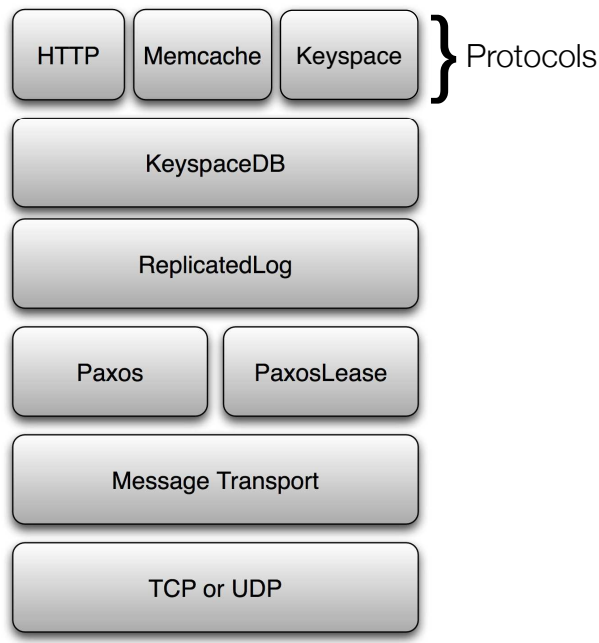}
\caption{Keyspace server program architecture.}
\end{center}
\end{figure}

In order to support both TCP and UDP-like network protocols, we implemented a \textbf{Message Transport} layer, which sits between Paxos and the actual network protocol-specific code. This allows us to use UDP for small messages and TCP for larger messages containing data.

Paxos, in the context of Keyspace is used to replicate database write commands (such as \texttt{SET, TEST-AND-SET} etc.). The Paxos algorithms only handles reaching consensus on a single value. A sequence of values are agreed upon by running different Paxos rounds sequentially. This functionality is encapsulated by the \textbf{ReplicatedLog} layer in Keyspace.

Once the ReplicatedLog finds that consensus has been reached on the next write command, it passes the command to the \textbf{KeyspaceDB} module. This is the module that actually stores the key-value pairs on local disk.

For local storage, Keyspace uses the Berkeley DB library's Transactional Data Store (TDS) engine. TDS provides the application (Keyspace in this case) with an ACID compliant data store. Keyspace uses TDS in a mode where the transaction log is flushed (synced) to disk on commit. This is an expensive operation which can take more than $10msec$; the consequence is that the bottleneck in Keyspace is the disk's ability to quickly flush data to the platter. The use of TDS is required by the distributed algorithm: in Paxos, nodes make promises to each other which they have to remember even after node restarts by storing them to stable storage.

Coming from the other side, clients can choose to connect using a variety of \textbf{protocols} such as HTTP and our own Keyspace protocol. To take advantage of all features and get maximum performance, users are advised to use the Keyspace protocol.

The protocol layer passes the client commands to the KeyspaceDB module, which, if it is a write operation and the node is the master, passes it to the ReplicatedLog, which will inject it into the underlying log sequence. Once the ReplicatedLog signals that it succeeded, KeyspaceDB will actually perform the write operations and return to the client.

If the client submitted a safe-read operation, the node will read the value(s) from its local database and return it. If the node is not the master it will only serve dirty-read requests. Clients can initially connect to any node to find out who the master is.

In terms of implementation, Keyspace is a completely asynchronous network server. On Linux, it uses the \texttt{epoll} event notification API, on BSD-like platforms it uses \texttt{kqueue}, while on Windows is uses the Completion Ports API. Since the Berkeley DB API is not asynchronous, we use multi-threading to avoid blocking the application during database operations.

\section{ Catchup }

Keyspace serves write requests if a majority of nodes are alive and connected. Nodes who are down or disconnected for some time and then rejoin the network need to \textit{catch up} to the rest. In Keyspace, there are two catchup-mechanisms:

\begin{enumerate}
\item If a node has been down or disconnected for a short time (ie. only few database commands have been entered into the replicated log), it can simply fetch the missing database commands from the up-to-date nodes who cache the tail of the log.
\item If a node has been down for a longer time (ie. the other nodes' cache is insufficient), the node will copy the master node's entire local database and then rejoin the Keyspace cell.
\end{enumerate}

The catchup mechanism is automated and requires no operator intervention. Nodes that are not up-to-date cannot obtain master leases.

\begin{figure}[htbp]
\begin{center}
\includegraphics[scale=0.5]{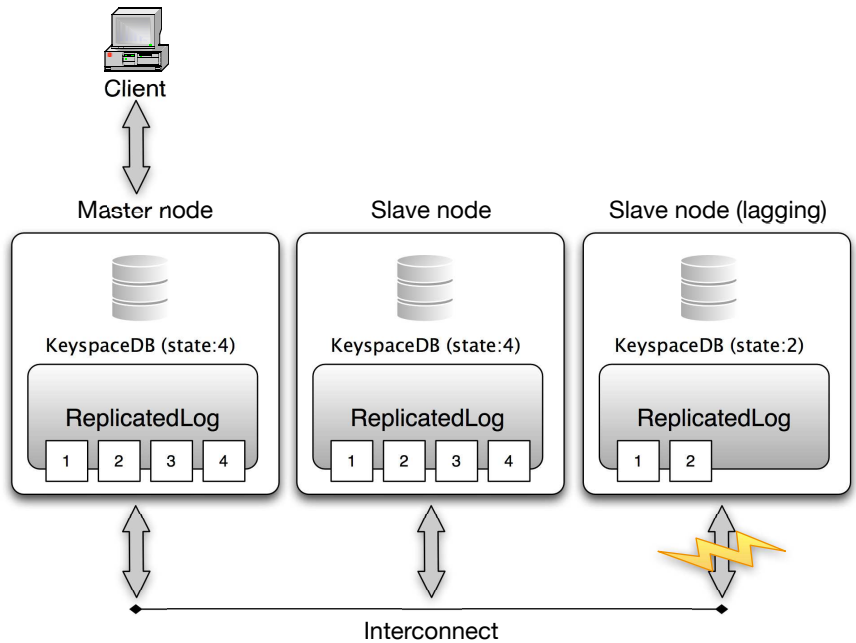}
\caption{Keyspace system with three nodes, one lagging behind. Two nodes have seen all four database commands, while one has seen only the first two.}
\end{center}
\end{figure}

\section{ Optimizations }

Keyspace contains a number of optimizations which reduce the number of network roundtrip times, reduce disk I/O and increase network throughput:

\begin{enumerate}
\item Command packing. In order to increase throughput, we pack several client commands into one Paxos round, up to 1M total. Since this is above the maximal UDP packet size of 64K, we use the TCP Message Transport system for Paxos.
\item Multipaxos. The classical Paxos algorithm requires two message roundtrips and two disk syncs (commits) per Paxos round. In master-based systems such as Keyspace, this can be reduced to one message roundtrip and one disk sync per round \cite{PaxosMadeLive}.
\item Commit chaining. Once a Paxos round is over, messages are passed up to the KeyspaceDB by the ReplicatedLog. KeyspaceDB then executes these commands on its local database. To avoid having to sync to disk (commit the transaction) at this point, these transactions are chained with the next Paxos round's and are only commited during the next Paxos round. Additional logic, outside the scope of this paper, ensures that this does not violate the strong consistency guarantees of Keyspace.
\end{enumerate}

\section{ Performance }

Keyspace was built from the ground-up as a high-performance network database and can service up to $\sim 100.000$ operations per second. Actual performance depends on data shape and access pattern, cache sizes, network speed and host hardware configuration such as CPU speed, main memory, and disk speed.

Figures 3 and 4 show Keyspace performance as seen by a single client performing reads and writes. The nodes are 2 x 2.4Ghz dual-core 64-bit AMD machines with 8GB of RAM and enterprise-class Western Digital SCSI hard disks (model WD4001ABYS), running 64-bit Debian GNU/Linux 5.0 with \texttt{reiserfs}. The tests were performed with the default Keyspace configuration, the relevant parameters being:

\begin{verbatim}
database.pageSize = 4096
database.cacheSize = 200M
database.logBufferSize = 10M
\end{verbatim}

Note that Keyspace uses master-slave replication where only the master serves safe reads --- but it can do so without contacting other nodes, since it is guaranteed that all prior write requests went through it. Hence there is no difference in \texttt{GET} throughput for single-server and replicated setups.

The difference between single-server and replicated \texttt{SET} thoughput is due to additional network roundtrips between nodes and dominated by individual disk sync times. Note that replicated write throughput levels off at $\sim$ 12MB/sec. Other thoughput values would drop off for value sizes (or key sizes) larger than $\sim$ 1000 bytes --- this is related to BerkeleyDB's page size and overflow logic. In general, for a database working with value sizes of $x$ bytes, \texttt{database.pageSize} should be set to $4x$.

\begin{figure}[h]
\begin{center}
\includegraphics[scale=0.5]{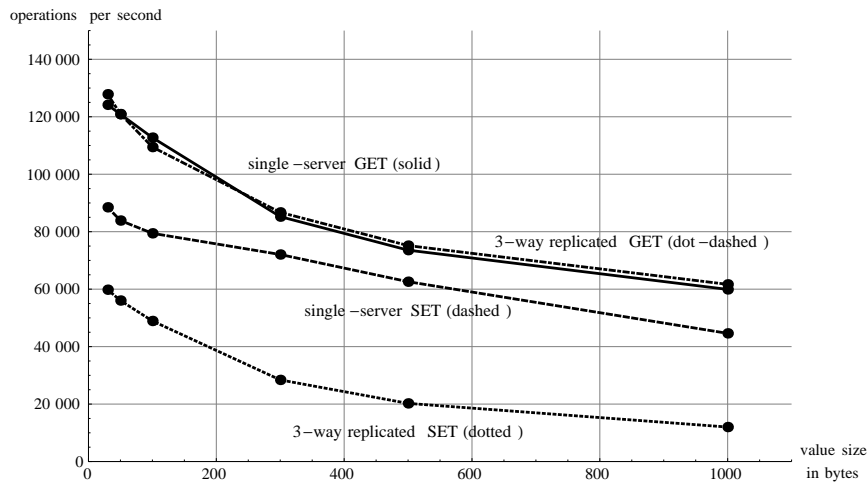}
\caption{Keyspace bulk throughput performance showing operations per second with key size = 30 bytes.}
\end{center}
\end{figure}

\begin{figure}[h]
\begin{center}
\includegraphics[scale=0.5]{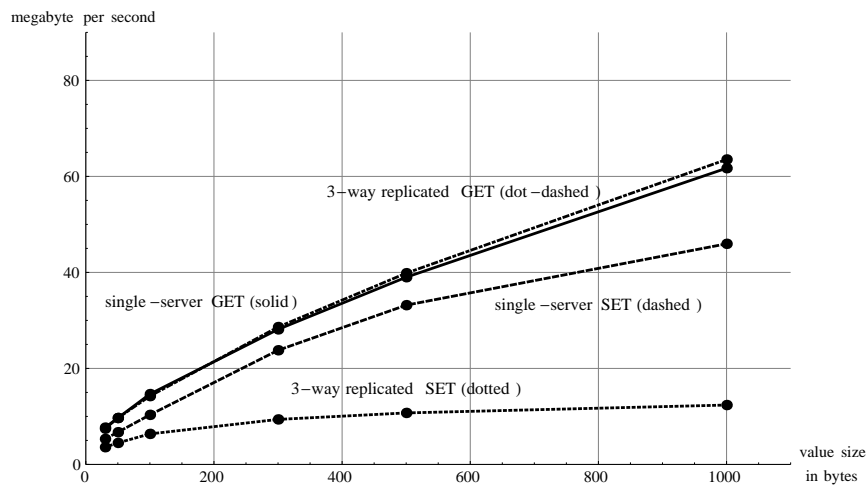}
\caption{Keyspace bulk throughput performance showing megabytes per second with key size = 30 bytes.}
\end{center}
\end{figure}

\section{ Further reading }

Leslie Lamport invented Paxos in 1990, but it was only published in 1998. This paper, \emph{The Part-Time Parliament} \cite{Parliament} was "greek to many readers", which led to a second paper, \emph{Paxos Made Simple}. Paxos was popularized by its use in Google's in-house distributed stack, described in \emph{Paxos Made Live - An Engineering Perspective} \cite{PaxosMadeLive} and \emph{The Chubby Lock Service for Loosely-Coupled Distributed Systems} \cite{Chubby}.


\begin{thebibliography}{9}

\bibitem{AGPL}
AGPL License, \url{http://www.fsf.org/licensing/licenses/agpl-3.0.html}.

\bibitem{PaxosMadeSimple}
L. Lamport, Paxos Made Simple, ACM SIGACT News 32, 4 (Dec. 2001), 18-25.

\bibitem{PaxosLease}
M. Trencseni, A. Gazso, PaxosLease: Diskless Paxos for Leases, 2009.

\bibitem{PaxosMadeLive}
T. Chandra, R. Griesemer, J. Redstone, Paxos Made Live - An Engineering Perspective, PODC '07: 26th ACM Symposium on Principles of Distributed Computing.

\bibitem{Parliament}
L. Lamport, The Part-Time Parliament, ACM Transactions on Computer Systems 16, 2 (May 1998), 133-169.

\bibitem{Chubby}
M. Burrows, The Chubby Lock Service for Loosely-Coupled Distributed Systems, OSDI'06: Seventh Symposium on Operating System Design and Implementation.

\end{thebibliography}
\end{document}